# Data-driven stability analysis in a multi-element supercritical Liquid Oxygen-methane combustor

Arijit Bhattacharya[1], Abhishek Sharma[1,2], Ashoke De[1*]

[1]*Department of Aerospace Engineering, Indian Institute of Technology Kanpur, 208016, Kanpur, India*

[2]*Liquid Propulsion Systems Center, ISRO, Valiamala, 695547, Thiruvananthapuram, India*

*Corresponding author: ashoke@iitk.ac.in

**Abstract**

Thermoacoustic instability (TAI) is a pressing problem in rocket combustors. TAI can cause significant damage to a combustor, resulting in mission failure. Therefore, stability analysis is crucial during the design and development phases of a rocket combustor. Stability analysis during the design phase can be substantially aided by the rocket combustor's large eddy simulation (LES). However, the computational cost of LES for full-scale rocket combustors is high. Therefore, using a small set of data from a large eddy simulation of a multi-element full-scale combustor, we investigated the effectiveness and computational needs of many data-driven and physics-driven tools for the classification of the stable and unstable regimes in the current study. Recurrence network analysis (RNA), reservoir computing (RC), and multi-scale permutation entropy (MPEA) analysis are the instruments employed in this study. The regime categorization task is unsuitable for RNA and MPEA, according to the results. With little input data, RC-based metrics may map the stable and unstable regimes and are thought to be computationally inexpensive and straightforward to use. In order to help with the design and development of rocket combustors, the combined LES-RC method to stability analysis is therefore anticipated to result in a notable decrease in processing needs.

## Introduction

Researchers have been interested in liquid rocket combustors used in launch vehicles for a number of decades [1–3]. Unfortunately, the nonlinear interactions between (i) injector and chamber, (ii) injector and feed system, (iii) numerous injectors, and (iv) many flames make them highly complex devices [4]. Rocket combustors are frequently studied using high-fidelity numerical simulations (e.g., 3D LES) as experimentation is complex and prohibitively costly. The liquid oxygen-methane ($LO_X$-methane) propellant technology is a promising technology for rocket propulsion due to its high specific impulse and smaller tank requirement [5]. Regretfully, the physics of the interaction between the various fuel injectors has been excluded from the majority of computational research on $LO_X$-methane combustors, which has concentrated on single-injector combustors [6–9].

Because of their high-pressure environment and transcritical propellant injection, supercritical rocket combustors are vulnerable to thermoacoustic instability [10]. Supercritical combustion dynamics have been numerically investigated in a number of recent publications [11–14], albeit using basic laboratory-scale combustor models. A few studies have concentrated on the combustion dynamics of supercritical LOX-methane combustors [15–17], although they only employed a single element (i.e., one fuel injector). According to earlier research, thermoacoustic instability can also be brought on by minor changes in rocket combustor operating parameters, such as a little drop in fuel injection temperature [3,4,14,18,19].

Thermoacoustic instability (TAI) is typified by high-amplitude pressure oscillation and extremely high heat release rates [20]. Heat release rate (HRR) oscillation and acoustic oscillation create a positive feedback loop that leads to TAI [21]. The positive connection between pressure and HRR can happen through a variety of mechanisms, such as abrupt changes in the equivalency ratio, oscillations in the surface area of the flame, and the formation of coherent structures in the flow [22]. During TAI, the rocket combustor is subjected to extreme pressure and heat flux levels, which frequently results in structural damage and combustor failure [23]. No reliable tool for early prediction of TAI exists to date [22]. Numerous studies have concentrated on the characterization and management of TAI in combustors as a result of this urgent issue. These can be roughly divided into three categories: (i) active control, (ii) stability analysis, which precisely maps stable and unstable operating ranges, and (iii) passive control. These are briefly described next to describe the completeness of the paper.

The efficacy of various physics-based techniques for early detection and active control of TAI has recently been explored. Complex networks [24–26], translational error [25], multifractal analysis [27], 0-1 test [28], cellular automata [29], and others are examples of these methods. A turbulence network model was employed by Hashimoto et al. [30] to identify thermoacoustic instability in a rocket combustor early on. Kobayashi et al. [16] employed a combination of k-means clustering, support vector machines, and ordinal partition transition networks to predict TAI early. They identified the crucial areas in the combustor driving the TAI using the transfer entropy metric. According to Krishnan et al. [31], as the system gets closer to thermoacoustic instability, significant power sources start to appear in the combustor. The high vorticity regions inside the combustor have a significant impact on the combustor dynamics, as demonstrated by another study [32] using complex network-based analysis.

A number of supervised machine-learning techniques have also been prescribed for use in active control systems. For instance, transfer learning [33], hidden Markov modeling [34], and symbolic time series analysis (STSA) [35–37]. TAI has lately been investigated using deep neural networks [38–40]. A 3D convolutional selective autoencoder was employed by Gangopadhyay et al. [39] to detect TAI early. In order to identify impending TAI, Ramanan et al. [41] employed the reconstruction loss found with a convolutional neural network-based variational auto-encoder (CNN-VAE) as an anomaly metric. Sengupta et al. [23] developed a robust early prediction framework for TAI using Bayesian neural network modeling with multimodal sensor data, such as temperature and pressure. A convolutional recurrent neural network (CRNN) was employed by Cellier et al. [42] to predict imminent TAI. A generative adversarial network technique was employed by Xu et al. [20] to detect TAI in solid rocket motors early. Regretfully, a lot of the deep learning algorithms in the literature only identify the dynamical transition to TAI when TAI has already started, which leaves little time for preventative measures [20]. Furthermore, because these models are frequently trained using data from a specific combustor, they have poor generalization. As a result, these methods are not appropriate for the early prediction of TAI.

A number of dynamic science-based tools have been explored for early prediction and active control of TAI. Many intriguing dynamical properties of the combustor during the transition to TAI have recently been discovered, including quasi-periodicity [43], classical intermittency [44], bursting mode oscillation [45], aperiodic intermittency-like behavior [26], etc. Hachijo et al. [46] collected early prediction metrics from the complexity-entropy causality plane (CECP) derived using pressure data from the combustor using hidden Markov modeling and k-means clustering. Lyu et al. [47] employed an artificial neural network and statistical indicators based on CECP to predict TAI early. However, before the CECP can be built using this method, the time series must be correctly embedded into the phase-space. Regretfully, human expertise is typically needed to accurately embed a time series in phase-space [48,49]. Therefore, it could be challenging to put the dynamical science-based techniques into practice.

On the other hand, stability analysis of combustors requires significant testing and prototyping [50], which are costly and time-consuming. Furthermore, the stability maps thus generated are specific to a particular combustor design and might not be very generalizable. Low-order nonlinear models are another method for stability analysis. Deb et al. [51] implemented fully convoluted network (FCN) and long short-term memory (LSTM) techniques, using data obtained from nonlinear reduced-order models, to early predict dynamical transition in complex natural systems. Matveev et al. [52] used a reduced-order model to produce a combustor stability map. Recently, Nóvoa and Magri [53] enhanced a low-order acoustic model's early TAI prediction capacity using a reservoir computing technique. However, low-order models may not generalize well because they are usually only applicable to the specific range for which they are designed [33].

The passive control strategy [54] lowers the propensity of the onset of TAI by redesigning burners or adding extra components inside the combustion chamber, such as vortex generators, Helmholtz resonators, baffles, acoustic liners, etc. Nevertheless, the passive control methods are relatively expensive.

We observe that the majority of research that has examined the transition to TAI using the physics-driven or data-driven approach has done so by analyzing experimental data after the combustor has already been constructed and developed. Therefore, this approach to combustor characterization permits only minimal combustor adjustments. In fact, relatively little research has been done on data-driven or physics-driven methods to support stability analysis in the design and development stage. In order to circumvent TAI, Chattopadhyay et al. [55] suggested a design approach based on STSA to alter the combustor geometry. However, this approach relies on experimental data generated from an already existing combustor. As a result, only minor design changes are possible in this situation. Designing the combustor optimally utilizing the information from prototype testing and experimentation [55] may be another strategy to prevent TAI. This would ensure that the combustor is not susceptible to TAI within the intended operating range. However, the cost of such a strategy is unaffordable.

Adopting high-fidelity CFD simulation (such as 3D LES) to investigate the combustor's stability range could be an alternate strategy. However, combustion modeling of rocket-scale combustors has primarily been conducted using single-injector element configurations. Although many numerical studies have investigated self-excited instability in gaseous oxygen (GOx)-methane single-element combustors, only a few have explored supercritical LOx-methane combustion. Using dynamic mode decomposition (DMD) and CFD data, Huang et al. [56] investigated key mechanisms supporting the TAI in a single-element rocket combustor. Zong et al. [57] utilized LES to investigate supercritical LOx-methane combustion in a single-shear coaxial injector with a laminar flamelet model. Urbano et al. [58] conducted LES on the BKD DLR combustor, simulating supercritical LOx-H$_2$ combustion. Recently, Sharma et al. [4,59] developed a supercritical LO$_X$-methane combustion modeling methodology simulating a seven-element swirl coaxial injector configuration. This model incorporates a real-gas equation of state to represent transcritical injection and employs a flamelet-generated manifold combustion closure to simulate combustion dynamics within the LES framework. They captured flame-flame and flame-wall interactions and elucidated the role of injectors in enhancing combustion dynamics. Furthermore, the LES framework was utilized to determine the impact of fuel injection temperature on the stability of a seven-element, rocket-scale combustor.

We note that combustor stability can be determined from LES-generated data itself. If the LES-generated pressure time series amplitude is low, then the combustor is considered stable. On the other hand, the high amplitude limit cycle oscillations in the time series typically indicate the onset of TAI. However, such an approach to stability mapping the combustor would require running LES simulations for a large set of operating points. Unfortunately, LES simulations require significant computational resources to correctly resolve the turbulence and chemistry [60]. Therefore, stability mapping of combustors with LES simulations alone is prohibitively costly. On the other hand, previous literature has shown that data-driven and physics-driven tools can accurately classify stable and unstable regimes with very few data points [24,26,61]. Therefore, it is essential to devise a suitable data-driven or physics-driven tool that can map the stability regime of a combustor using the LES-generated time series with a few data points. This approach will require a lower number of LES simulations, resulting in significant savings in computational resources. Because stability analysis is a critical issue in combustor design, suitable data-driven or physics-driven stability analysis of CFD data can significantly aid the design and development of combustors. Unfortunately, to the best of the author's knowledge, there are no studies in which this approach to stability analysis is taken. To bridge this literature gap, we study the efficacies of three well-known data-driven and physics-driven tools in stability mapping of the LO$_X$-methane combustor, namely multi-scale permutation entropy analysis, recurrence network analysis, and reservoir computing. The pressure time series produced by a 3D LES simulation of the LO$_X$-methane combustor [4] is used as the input data for this study. The findings demonstrate that reservoir computing is a promising method for mapping the combustor's stability regime using a small number of data points.

**Brief theoretical background of the data-driven tools used in the present study**

In the present study, we explore the dynamical transition to thermoacoustic instability (TAI) using three well-known data-driven or physics-driven tools, namely the multi-scale permutation entropy analysis (MPEA), recurrence network analysis (RNA), and reservoir computing (RC), which are briefly described below.

*(a) Multi-scale permutation entropy analysis*

Entropy and related measures from a time series indicate the regularity (i.e., orderliness or determinism) of the time series [62]. Entropy is high for a disorderly system and is low for an orderly or deterministic system. A commonly used entropy is the permutation entropy [63].

For a discrete time series $X = \sum_i x_i = \{x_1, x_2, \cdots, x_n\}$, the measure permutation entropy $H_X$ indicates its mean uncertainty. It is given by,

$$H(X) = -\sum_{x_i \in \theta} p(x_i) \ln p(x_i)$$

Here, $\theta$ includes all the values that $X$ may take. $p(x_i)$ is the associated probability that $X$ takes the value $x_i$. Unfortunately, permutation entropy can capture single-scale data only [62]. Recently, multi-scale permutation entropy (MPE), a variant of the permutation entropy, has been suggested by Costa et al. [62]. MPE can capture multi-scale data and is more robust against noisy signals than traditional permutation entropy [62,64]. MPE has been previously used for early detection of lean blowout combustion instability [65].

Now, let us discuss the concept of multi-scale permutation entropy. Let a time series be given as $\{x_1, x_2, \cdots, x_n\}$. From it, a coarse-grained time series is calculated. Let an element of the coarse-grained time series is denoted by $y_j^{(s)}$ where $s$ is the scale factor, and $j$ is constrained by, $1 \leq j \leq L/s$. Then, $y_j^{(s)}$ is given as, $y_j^{(s)} = 1/s \sum_{i=(j-1)s+1}^{js} x_i$. If $s = 1$, the coarse-grained time series becomes identical to the original time series. Let $m$ be the number of values (states) possible for the time series. Then, at most, there could be $m!$ number of permutations (patterns) of the time series elements. Let us consider that the $i^{\text{th}}$ pattern is denoted as $\pi_i$. Then, the multi-scale permutation entropy (MPE) is given by,

$$\text{MPE} = -\sum_{i=1}^{m!} p(\pi_i) \log p(\pi_i)$$

Here, $p(\pi_i)$ is the probability of pattern $\pi_i$ to occur. The, normalized multi-scale permutation entropy (NMPE) is given by,

$$\text{NMPE} = \text{MPE}/\log m!$$

A large value of $m$ is needed to correctly capture all the different states in the time series. Because NM$PE$ is a type of entropy, it can be said that a high value of NM$PE$ indicates disorderliness. The maximum value of NM$PE$ is 1, when all permutations have equal probability [66]. On the other hand, a low NM$PE$ indicates a high orderliness. The minimum value of NM$PE$ is 0, when the time series is highly regular or highly deterministic (e.g., highly periodic) [66].

*(b) Recurrence network analysis*

Recurrence network analysis (RNA) is the second technique employed in this study to investigate the dynamical transition. Any dynamical system that frequently returns to its original state has recurrence as a fundamental characteristic [49]. The recurrence plot (RP) is a symmetric time vs. time plot that graphically depicts the recurrence information of the underlying dynamical system. There is just binary data in the RP. If at two time instances the dynamical system recurs (i.e., comes into close proximity in phase space), then the corresponding point in the RP is marked black. In the absence of a recurrence, the corresponding point is marked white. The recurrence network (RN) is a complex network based on the RP. As such, it is a complementary approach to early detection and the characterization of thermoacoustic instability [24,26,67].

Next, we briefly discuss the calculation of RN and the RN-based metrics for early detection of TAI. First, the time series obtained from CFD simulation is embedded in phase space using Taken's embedding theorem [68]. The optimal embedding dimension ($M$) is calculated using the false nearest-neighbor method [69]. The optimal time delay is the time delay when the autocorrelation function first crosses zero [70]. The phase-space vectors thus generated are considered to be nodes of the recurrence network. Next, $M$-dimensional hyperspheres are drawn around the phase-space vectors using a recurrence threshold $\epsilon$. $\epsilon$ is calculated to be 10% for the present study using the method of Eroglu et al. [71]. If two hyperspheres overlap with each other, then the corresponding nodes are said to be connected via a link in the recurrence network. From the RN thus constructed, three RN metrics, namely the global efficiency ($\eta$), average degree centrality ($k$), and global clustering coefficient ($C_G$) are calculated. More details about the RN approach and the physical significance of the RN metrics can be found in the appendix of this paper as well as in previous studies [26,72].

*(c) Reservoir computing (RC)*

Reservoir computing, a kind of recurrent neural network (RNN), is the third technique used in this investigation. A popular computational method based on the workings of a biological brain is the recurrent neural network (RNN) [73]. An RNN is a collection of many neurons, which are processing units. Neurons send feedback signals to one another and are intricately linked. RNNs have a number of benefits when it comes to simulating dynamical systems. They have a dynamic memory and somewhat mimic how the brain functions naturally. The error backpropagation (BP) algorithm, which modifies the network weights by minimizing the error, is used to train RNNs. RNN training, however, requires a lot of resources [73].

Both Jaeger [74] and Maass et al. [75] independently introduced a unique kind of RNN-based method that is now called reservoir computing (RC). RC and deep learning (DL) are very similar. Both of these machine-learning techniques are based on RNNs and match input data with target output data (ground truth) [73]. However, when it comes to the RNN's training process, DL and RC differ significantly. Both the output weights and the RNN weights are trained in DL. In contrast, just the output weights are taught in RC, while the RNN (also known as the reservoir) is generated at random. Throughout the training process, the RNN itself stays constant. A linear combination of the signals derived from the neurons in the reservoir is then used to reproduce the target output. Because of the difference in training, RC has a significantly smaller parameter size than DL, which leads to quick training and minimal computational demands [76]. Prior research has demonstrated that RC performs exceptionally well in modeling dynamical systems despite the smaller parameter size [77–82]. The RC technique is schematically explained in Fig. 1.

Because of its benefits, RC has been used in a number of domains, such as nonlinear dynamics [83], physical computing [84], and cavitation detection [85]. The RC technique has recently been applied in numerous thermo-fluid science studies. Pandey and Schumacher [86] modeled a 2D turbulent Rayleigh-Bénard convection flow using reservoir computing and discovered that RC could accurately capture the underlying physics. Using the RC technique, Valori et al.

[79] recreated severe vorticity occurrences in turbulent Rayleigh-Bénard convection flow. The RC technique was utilized by Yao et al. [80] to learn different ocean circulation models. Nevertheless, the RC approach has not been applied in any of the research mentioned above to capture dynamical transition or for regime classification task. Instead, the majority of these studies concentrated on making precise predictions about the future time series trajectory of the relevant systems.

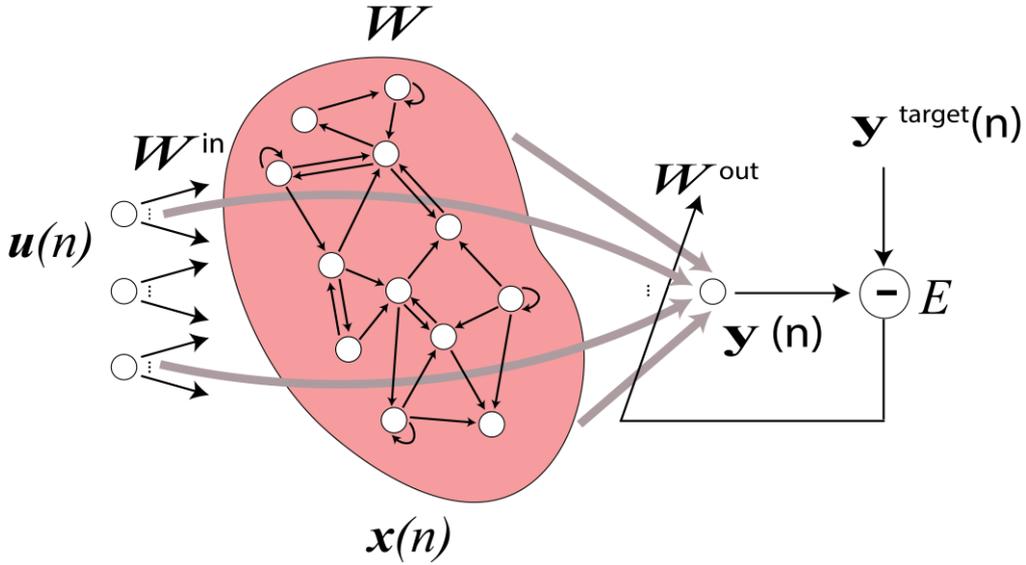

Fig. 1: Schematic representation of a RC network. $\boldsymbol{u}(n)$ is the input, $\boldsymbol{W}^{\text{in}}$ indicates the input weights, $\boldsymbol{W}^{\text{out}}$ indicates the output weights. $\boldsymbol{x}(n)$ is the reservoir state, $n$ is the time instant, $\boldsymbol{y}(n)$ is the output produced by the RC while $\boldsymbol{y}^{\text{target}}(n)$ is the target output (ground truth). $E$ indicates the error between RC output and ground truth.

Quantifiable metrics are required if a regime classification method for stability analysis is to be constructed utilizing the RC framework. To determine the prediction error of RC, Liang et al. [77] employed statistical measures. However, rather than using those for regime categorization, they utilized them to investigate how well the RC learned the time series trajectory. As far as the author is aware, this work is the first to create measurable metrics based on RC for classifying regimes and identifying dynamical shifts in a system.

Next, we describe the mathematics behind reservoir computing. The reservoir in the RC is a (i) high-dimensional expansion of input time series as well as (ii) a dynamic memory of the input data [73]. The data coming from complex systems like a turbulent combustor is unlikely to be linearly separable. Therefore, the original data are expanded to higher dimensions to obtain the reservoir states, which are often linearly separable [73]. In the present study, RC is trained using a numerically obtained pressure time series of duration 0.06 seconds with a sampling frequency of 0.1 MHz. Every time series is divided into two parts: the first 80% of the time series is used for training the output weights, and the last 20% is for validation (future time series trajectory prediction). Let the training time series be $\boldsymbol{u}(n)$. Further, consider that the output time series is known and given as $\boldsymbol{y}_{\text{target}}(n)$ where, $n$ is the time instant given by $n =$ $1, 2, \ldots, T$. Here, $T$ is the number of data points in the training dataset. Therefore, the training dataset is indicated by, $\{(\boldsymbol{u}(n), \boldsymbol{y}_{\text{target}}(n))\}$.

Let $N_{in}$ and $N_{out}$ be the number of the input and output channels. In the present study, only one input time series is fed into the reservoir, and one output time series is obtained from the reservoir, i.e., $N_{in} = N_{out} = 1$. Next, we discuss the creation of the recurrent weight matrix $\boldsymbol{W}$, which is the adjacency matrix of the reservoir. In the present study, the size of the reservoir is kept constant at $N=200$ to reduce the computational time. A matrix $\boldsymbol{W}'$ of size, $N \times N$ is randomly generated based on Gaussian distribution. The maximal absolute eigenvalue of the matrix $\boldsymbol{W}'$ is calculated. This variable is called the spectral radius $\rho(\boldsymbol{W}')$ of the reservoir. Next, $\boldsymbol{W}'$ is divided by $\rho(\boldsymbol{W}')$ to get the recurrent weight matrix $\boldsymbol{W}$ having a unit spectral radius.

The next task is the creation of the input weight matrix $\boldsymbol{W}^{\text{in}}$. $\boldsymbol{W}^{\text{in}}$ is a matrix of size $N \times N_{in}$ and also is generated randomly based on Gaussian distribution. We note that the seed to create the random Gaussian distributions is kept fixed at 42 so that, in all cases, the reservoir weights and input weights remain the same. Let the high-dimensional reservoir state be given by the vector $\boldsymbol{x}(n)$. $\boldsymbol{x}(n)$ is initialized as a zero element matrix of size $N$ and then updated for every $n$ as follows:

$$\boldsymbol{x}(n) = (1-\alpha)\boldsymbol{x}(n-1) + \alpha\tilde{\boldsymbol{x}}(n) \tag{1}$$

Here, $\alpha \in (0,1]$ is the constant leaking rate. We arbitrarily consider $\alpha = 0.3$. $\tilde{\boldsymbol{x}}(n)$ is the update of the reservoir state and is given by,

$$\tilde{\boldsymbol{x}}(n) = \tanh(\boldsymbol{W}^{\text{in}}[1; \boldsymbol{u}(n)] + \boldsymbol{W}\boldsymbol{x}(n-1)) \tag{2}$$

Here, $\tanh(\blacksquare)$ is applied elementwise and; indicates vertical concatenation (i.e., matrix concatenation). Next, the optimal output weights $\boldsymbol{W}^{\text{out}}$ that minimize the squared error between the RC output $\boldsymbol{y}(n)$ and the ground truth $\boldsymbol{y}_{target}(n)$ are found out. This amounts to solving a typically overdetermined system of linear equations given by [73],

$$\boldsymbol{y}^{\text{target}} = \boldsymbol{W}^{\text{out}} \boldsymbol{X} \tag{3}$$

Here, $\boldsymbol{X} = [1; \boldsymbol{u}(n); \boldsymbol{x}(n)])$. The most general and stable solution to (3) is ridge regression, also known as regression with Tikhonov regularization. This amounts to solving the following equation,

$$\boldsymbol{W}^{\text{out}} = \boldsymbol{y}^{\text{target}} \boldsymbol{X}^{\text{T}} (\boldsymbol{X}\boldsymbol{X}^{\text{T}} + \beta \boldsymbol{I})^{-1} \tag{4}$$

Here, $\boldsymbol{I}$ is the identity matrix, and $\beta$ is a regularization coefficient. We take $\beta = 1e^{-8}$ similar to previous studies in reservoir computing [82]. Next, the RC output is given as, $\boldsymbol{y}(n) = \boldsymbol{W}^{\text{out}}[1; \boldsymbol{u}(n); \boldsymbol{x}(n)]$. Here, $\boldsymbol{W}^{\text{out}}$ is the optimal output weight.

We note that the various global parameters (or hyper-parameters) may be optimized to improve the performance of the reservoir [73]. However, our intent in the present study is only to show the proof of concept that reservoir computing can be used for regime classification in a TAI-prone combustor. Therefore, we do not consider the optimization problem of the hyper-parameters in the present study.

Next, the trained reservoir is used to predict the remaining 20% of the time series. Three error metrics based on RC are calculated as follows. Normalized mean square error ($NMSE$) is given by,

$$NMSE = \frac{\sum_{i=1}^{n}(\overline{y_m}-y_i)^2}{\sum_{i=1}^{n}(\overline{y_i}-y_i)^2} \tag{5}$$

Here, $y_i$ is the mean of the real signal. $\overline{y_m}$ is the mean of the predicted signal. A second metric, mean absolute error ($MAE$), is given by,

$$MAE = \sum_{i=1}^{n} \frac{\overline{y_m} - y_i}{n} \qquad (6)$$

The third RC metric, Root mean square error ($RMSE$), is given by,

$$RMSE = \sqrt{\sum_{i=1}^{n} \frac{(\overline{y_m} - y_i)^2}{n}} \qquad (7)$$

One way to implement RC is to train the reservoir for each operating point to minimize the prediction error. This approach to RC may be useful when learning the time series trajectory itself. However, if RC is to be used to capture a dynamical transition in a given system, such an approach will lead to significant complexity. In such a case, the reservoir has to be optimized for every possible dynamical state. This will require adjustment of the various hyperparameters of the reservoir ($N, \alpha, \beta$, etc). Such adjustments, however, require input from a human expert [73] and, therefore, cannot be easily implemented. An alternative way may be to keep the reservoir fixed while training the output weights using the training dataset and then calculate the prediction errors for the various dynamical states. With such an approach, we can track the changes in system dynamics by monitoring the RC metrics, as described next.

As discussed above, the RC metrics $NMSE, MAE$, and $RMSE$ represent the error in predicting the future time series trajectory. Previous studies have reported that during the transition to TAI, the dynamics shift from chaos (at a stable combustion state) to highly deterministic (at TAI) [26,87]. It is well-known that a time series can be predicted with reasonable accuracy only if it is deterministic [88]. On the other hand, chaos, by definition, is unpredictable. Therefore, it is likely that the RC metrics, which quantify the error in the prediction of future time series trajectory, will change from a high value at the stable combustion state to a low value at TAI. We track the changes in the RC metrics during the dynamical transition to see whether they follow a consistent trend. As the RC metrics thus evaluate the determinism in the time series, we infer that they have a sound physical basis and, therefore, are likely to be applicable to combustors showing similar dynamical transition to TAI with minimum calibration. Moreover, the RC approach devised in this paper is easy to implement and also computationally cheap, as will be shown later.

**CFD framework**

Here, we provide a brief introduction to the CFD framework used to generate the pressure time series. Finer details on the CFD numerical framework can be refereed from previous works by Sharma et al. [4,59]. In this framework, large eddy simulation (LES) is used to capture turbulent mixing, heat release rate oscillations, and acoustic wave transmission leading to combustion instability. Multiple large eddy simulations are carried out using the finite-volume solver Ansys Fluent. The nonlinear thermodynamic and transport property variations due to transcritical and supercritical regimes are captured using the Soave-Redlich-Kwong (SRK) and Chung transport properties [89]. We use Chapman–Enskog theory with dense fluid corrections. This LES framework, incorporating real gas thermodynamics and a high-fidelity flamelet-generated manifold combustion model, is utilized to accurately resolve turbulence and combustion scales, capturing the effect of low fuel injection temperature on the stability of the LO$_X$-methane combustor. The framework was validated against the benchmark Mascotte G2 test case from ONERA, which operates on LO$_X$-methane propellants. The numerical methodology accurately reproduced transcritical oxygen injection and the supercritical flame structure. The validated methodology is then applied to the rocket-scale, seven-element bi-directional swirl coaxial injector combustor, as shown in Fig. 2, operating on LO$_X$-methane propellants. The methodology was seen to effectively capture the multi-element rocket combustor flow, flame interactions, and significant flow features. A comprehensive dataset of pressure statistics by conducting multiple LES at various fuel injection temperatures was generated and is currently being utilized in this study.

*Computational domain*

In the present study, a full-scale combustor with seven bi-directional swirl coaxial injectors is chosen as the computational domain (Fig. 2). The inset image shows the cutaway view of an injector and input paths of methane and liquid oxygen propellants. Oxygen is input into the combustor at a transcritical state with an injection pressure of 70 bar and temperature of 83K. On the other hand, methane injection temperature varied from 233 K to 200 K. Mass flow inlet boundary condition is applied at both fuel and air injectors. Combustor walls are considered adiabatic, and no-slip conditions are imposed. More details about the combustor geometry may be found in our previous study [4,59].

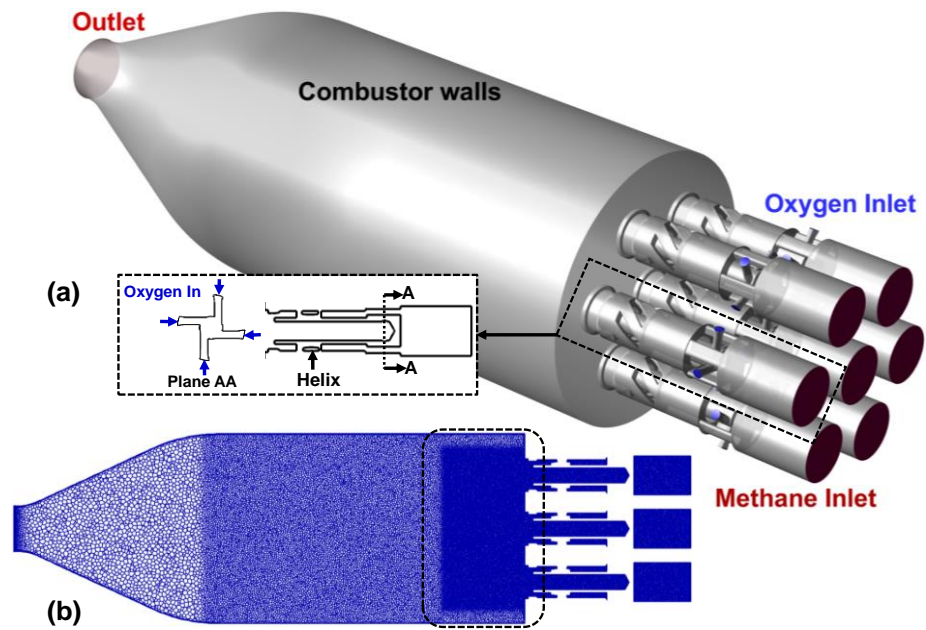

Fig. 2: Computational domain- Multi-Element Combustor. Reproduced from. Sharma *et al.*, Physics of Fluids 37, 025222 (2025). Copyright 2025 AIP Publishing, LLC.

An adequately refined unstructured mesh is created around the injectors, which resolves more than 80% of turbulent scales. A sufficiently small time step is chosen to capture the unsteady dynamics. To minimize the effects of initial transients, each case is run for 40 flow-through cycles, corresponding to more than 300 cycles for the first longitudinal (1L) mode and 600 cycles for the first transverse (1T) mode, respectively. Next, the case was further run for 0.06 seconds at a sampling rate of 0.1 MHz, generating a pressure time series of 6000 data points.

Fig. 3 is a cut section of the combustor showing the pressure probes located in the combustor. The injector probes, chamber probes, and methane inlet probes are respectively indicated as IP, CP, and IM. In the present study, the pressure time series obtained at probe CP1 is used in the time series analysis.

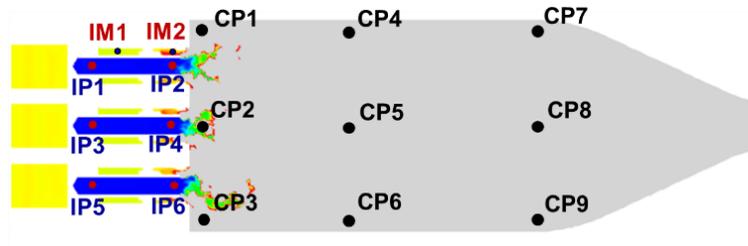

Fig. 3: Pressure probe location in combustor. Reproduced from. Sharma *et al.*, Physics of Fluids 37, 025222 (2025). Copyright 2025 AIP Publishing, LLC.

**Result and discussion**

The present study uses the pressure time series at various dynamical states obtained with our previous numerical study [4,59]. The validation of the numerical model used to generate the pressure time series is provided in our previous study titled "Numerical investigation of supercritical combustion dynamics in a multi-element LOx–methane combustor using flamelet-generated manifold approach [4].

Fig. 4 shows the numerically obtained pressure time series $p$ for various fuel injection temperatures. The pressure time series shows low-amplitude, seemingly random oscillations at a fuel injection temperature of 233K (Fig. 4(a)). Combustion noise, or the stable combustion condition, is characterized by such a circumstance. The oscillations' amplitude considerably rises when the fuel injection temperature is lowered to 219K (Fig. 4(b)). Furthermore, the zoomed-in time series shows that some periodicity has been added to the time series. The time series is still aperiodic generally, though. The pressure time series becomes almost periodic (i.e., extremely deterministic) when the fuel injection temperature is slightly lowered to 217K. Additionally, the oscillations' amplitude grows (Fig. 4(c)). These facts indicate the onset of thermoacoustic instability. The peak-to-peak amplitude of oscillations increases even more as the fuel injection temperature is lowered to 200K, indicating that thermoacoustic instability becomes more prominent (Fig. 4(d)). However, even at the TAI state, there are notable variations in the peak-to-peak amplitude of oscillations, as seen by close examination of time series at TAI (Fig. 4(c-d)).

Fig. 5 shows the power spectra obtained from the pressure time series $p$ at various temperatures. At the stable combustion state (Fig. 5(a)), no prominent peak of significant amplitude is seen. On the other hand, for a fuel injection temperature of 200K, a high-amplitude distinct peak at 4346 Hz is seen in the power spectrum (Fig. 5(b)). This indicates that the corresponding state is indeed TAI.

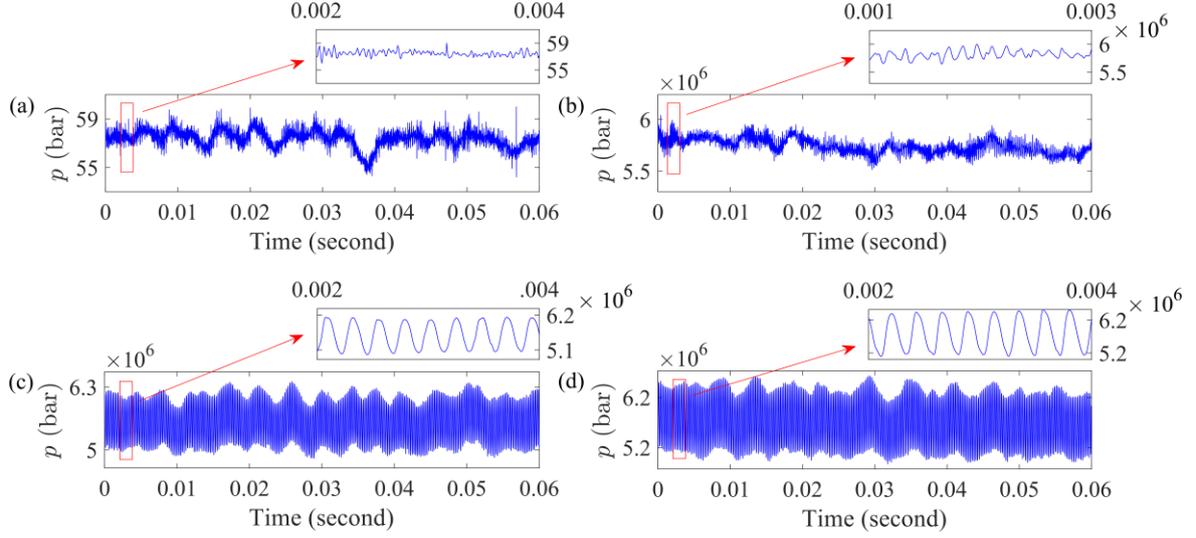

Fig. 4: Numerically obtained pressure time series $p$ at various fuel injection temperatures. (a) 233K, (b) 219K, (c) 217K, and (d) 200K.

*(a) Multi-scale permutation entropy analysis*

We then go into how well the NMPE metric captures the dynamical transition from the stable combustion state to TAI (Fig. 6). We observe that NMPE is substantially lower at the TAI state (200K) than it is at the stable combustion state (233K). This indicates highly disorderly (i.e., chaotic) dynamics at the stable combustion state. TAI state dynamics, on the other hand, are more regular (i.e., periodic). Therefore, the observations made from the time series visualizations are also corroborated by the normalized multi-scale permutation entropy (NMPE). Up until 219K, the NMPE acquired using scale-1, or the original time series, exhibits a little declining trend. Then, at 217K, when TAI starts, it suddenly drops to a very low value. On the other hand, NMPE obtained with the other scales initially remains almost constant as the temperature is reduced but abruptly reduces when TAI is onset (at 217K). We conclude that NMPE is an inappropriate metric for early identification of imminent TAI for the dynamical transition seen in the current investigation.

A significant drawback of the multi-scale permutation entropy is that it does not take into account the amplitude variation information when searching for ordinal patterns (i.e., ordering patterns) [90]. In practice, the ordinal patterns resulting from small fluctuations in amplitude may be due to the inherent noise in the combustor dynamics and, therefore, should have a lower contribution to the calculated NMPE metric [90]. On the other hand, the ordinal patterns resulting from higher amplitude fluctuations are likely to be indicative of the predominant dynamics of the system and, henceforth, should have a higher contribution. Because the present system does have significant amplitude oscillations at the TAI, this drawback causes the NMPE metric to fail to capture the transition to TAI.

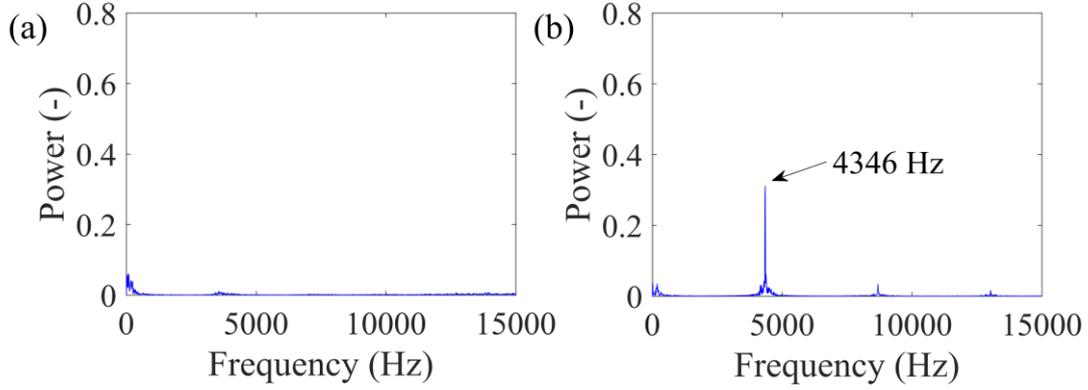

Fig. 5: Power spectra obtained from the pressure time series at various fuel injection temperatures. (a) stable combustion state (input fuel temperature of 233K) and (b) thermoacoustic instability state (input fuel temperature of 200K).

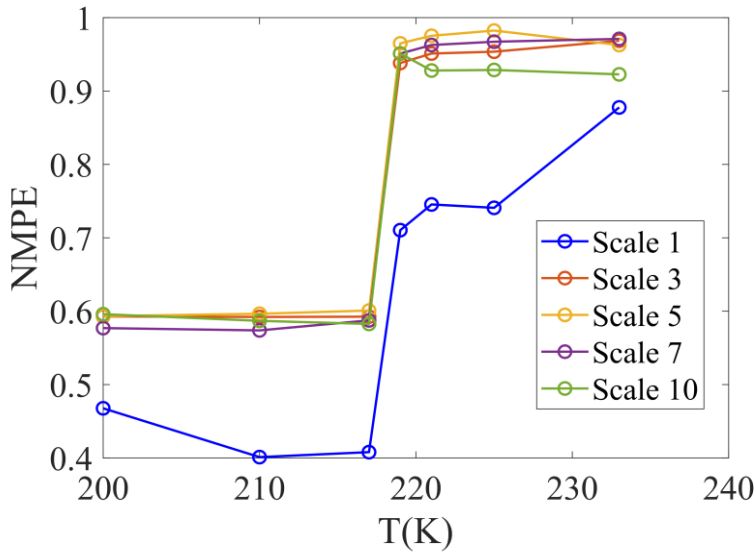

Fig. 6: Variation of NMPE as the fuel injection temperature is varied to approach TAI.

*(b) Recurrence network analysis*

Fig. 7 shows the recurrence plots (RP) corresponding to the stable combustion state (Fig. 5(a)) and the TAI state (Fig. 7(b)). In Fig. 7(a), the grainy structures present in the RP consist of very short horizontal and diagonal lines. Such structures are indicative of chaotic dynamics [44]. On the other hand, the RP at the TAI state (Fig. 7(b)) comprises short diagonal lines and white patches in between them. The diagonal lines have almost constant gaps between them, which is representative of periodic dynamics. The white patches are indicative of dynamical states that are much different from the normal state of the system [49], which is likely caused by the amplitude modulation in the signal. The amplitude modulation of the time series also results in a lack of long diagonal lines in the RP. Therefore, RP-based approaches, e.g., recurrence quantification analysis, are unlikely to capture the accurate dynamics at TAI.

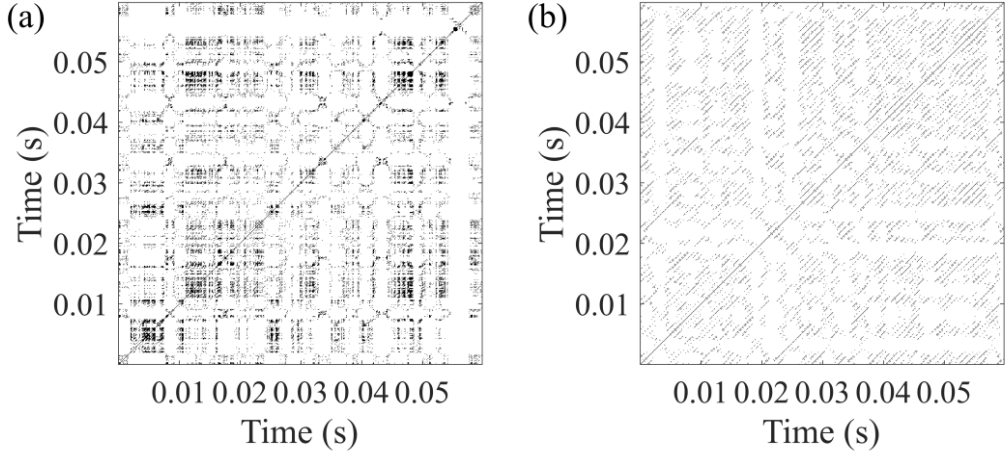

Fig. 7: Recurrence plots corresponding to various input fuel temperatures. (a) 233K and (b) 200K.

Prior research has demonstrated that the dynamical transition from chaotic dynamics to periodicity can be precisely captured by recurrence network (RN) analysis [24,26,67]. RN is a complex network created by removing the temporal information from the time series and only considering the amplitude information [91]. As such, it is a complementary approach to RP-based dynamical analysis. We then use RN to investigate the dynamical transition. Fig. 8 shows the monotonically falling trend of the RN measure $\eta$ as TAI approaches. It is known that $\eta$ is higher if the dynamical state is deterministic (i.e. periodic) and is lower if the dynamical state is chaotic [26]. Therefore, the results indicate that determinism decreases with the approach of TAI. However, it is evident from a visual examination of the time series (Fig. 4) that this is not the case. On the other hand, as TAI approaches, the RN measures $k$ and $C_G$ do not exhibit any discernible trends. We conclude from the explanation above that RN is not a good early prediction tool for the dynamical shift to TAI that is occurring here.

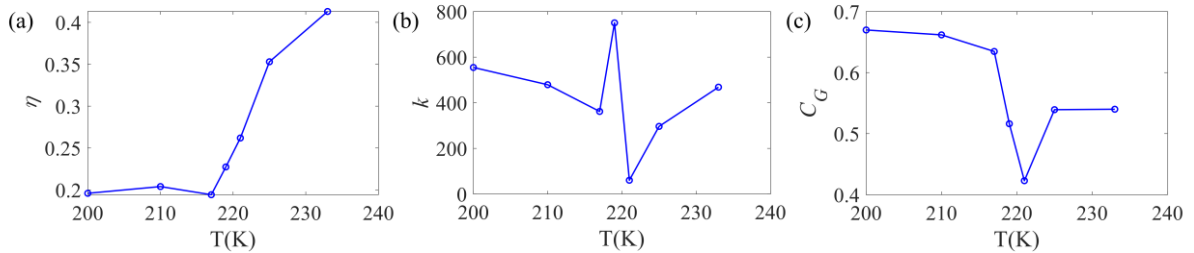

Fig. 8: Variation of the RN metrics $\eta, k$ and $C_G$ as the fuel injection temperature varies when approaching TAI.

The failure of RNA to capture the transition towards TAI in the present study can be easily explained. We recall that RNA finds the patterns in the data using a fixed recurrence threshold.

This causes a loss of important information about the dynamics of the system contained in the combustor and may cause RNA to fail to capture the dynamical transition [92]. Moreover, as discussed above, RNA is based on amplitude correlations embedded in the data. Therefore, the significant amplitude fluctuations observed at the TAI and near-TAI states may have caused a substantial loss of the amplitude correlations. This might have failed RNA in capturing the dynamical transition to TAI. This fact can also be corroborated by the recurrence plot at the TAI state (Fig. 7(b)), where the long diagonal lines typically seen at TAI are not present. It may be noted that the previous studies [24,26,67], which implemented RNA to successfully characterize/early-predict TAI, did not find significant oscillations in the time series amplitude at TAI, unlike the present study.

*(c) Reservoir computing approach*

Next, the reservoir computing approach is used to investigate the dynamical transition. It should be noted that prior to being fed into the RC, the pressure time series $p$ is standardized. The formula for the standardized pressure time series is $P = (p - m)/SD$. The pressure time series $p$ has a mean of $m$ and a standard deviation of $SD$. As discussed previously, the output weights of the RC are trained using the first 80% of the time series. The remaining 20% of the target time series (i.e., the ground truth) and corresponding RC-predicted time series for the different fuel injection temperatures are displayed in Fig. 9. The RC cannot accurately forecast the future time series trajectory at the stable combustion state (Fig. 9(a)). A reasonably accurate forecast can be made for a fuel injection temperature of 219K (Fig. 9(b)). A significantly high prediction accuracy is achieved for the fuel injection temperature of 217K, when TAI is onset (Fig. 9(c)). For the 200K fuel injection temperature, where the TAI is more noticeable, an even better forecast is made (Fig. 9(d)).

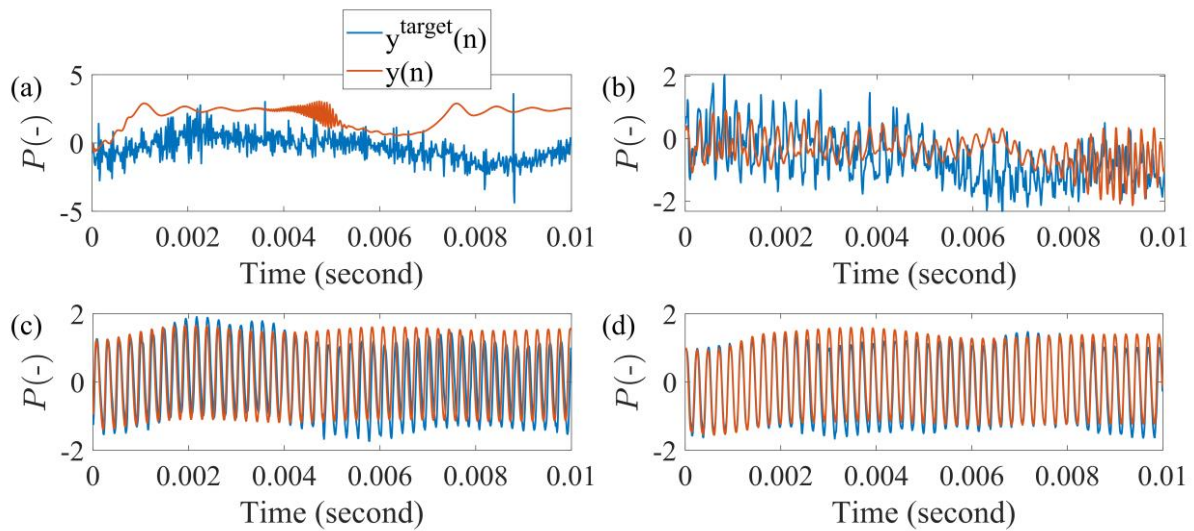

Fig. 9. Predicted and target time series for various fuel injection temperatures (a) 233K, (b) 219K, (c) 217K, and (d) 200K. Note that the pressure time series are standardized by, $P = (p - m)/SD$. Here, $m$ is the mean and $SD$ is the standard deviation of the pressure time series $p$. For ease of visualization, 0.01 second long time series is shown here (corresponding to 1000 data points).

The varying accuracy of the RC model in predicting the future time series trajectory, as seen above, is easily explained. At high injection temperatures, the pressure time series consists of apparently random low-amplitude oscillations around the mean (Fig. 4(a)). Such a time series is indicative of chaotic dynamics. It is known that for chaotic systems, the true and RC-predicted time series gradually diverge with time [78], as the deterministic RC cannot accurately predict the chaotic time series well for extended periods. Indeed, chaos by nature is unpredictable [93]. Therefore, at high fuel injection temperatures, significant deviations are expected between the RC-predicted time series and the numerically obtained pressure time series. On the other hand, the RC model accurately predicts the pressure time series at lower fuel injection temperatures, which consist of highly periodic oscillations (Fig. 4(c-d)). Future time series of a periodic signal can be quite accurately predicted if the prediction interval is known. This is because periodic dynamics recur in time [49]. Indeed, previous studies have shown RC to be able to accurately predict future time series trajectories of highly periodic signals [77]. Therefore, at low fuel injection temperature, the RC-predicted time series and the numerically obtained pressure time series match quite closely.

At different fuel injection temperatures, Fig. 10 displays the time series prediction error as measured by the RC metrics. At first, we observe that when the fuel injection temperature is lowered from 233K (stable combustion state), $NMSE, MAE$, and $RSME$ all follow declining patterns. However, at around 219K, the metrics $NMSE$ and $MAE$ become nearly constant and then gradually decline from there. On the other hand, after reaching a low value at 219K, $RSME$ initially stays nearly constant before again continuing to decline. At 219K itself, it is observed that $NMSE, MAE$, and $RSME$ achieve nearly identical values as those at 217K (i.e., at the commencement of TAI). To correctly distinguish stable and unstable regimes, we set the following limits for the three RC metrics: $NMSE \leq 1.5, MAE \leq 0.32$, and $RSME \leq 0.9$. The dynamical states corresponding to the fuel injection temperature of 219K to 200K are then shown to belong to the unstable combustion regime, according to the RC. We remember that the TAI does not start before the 217K fuel injection temperature. The RC technique automatically adds a safety margin to the stability map by identifying the combustor as unstable at 219K. In aeronautical vehicles, a safety margin in the combustor stability map is ideal. This is because there are a number of inherent uncertainties in the operation of aerospace vehicles [94]. Such uncertainties are addressed by incorporating a safety buffer, which results in a robust vehicle design [94].

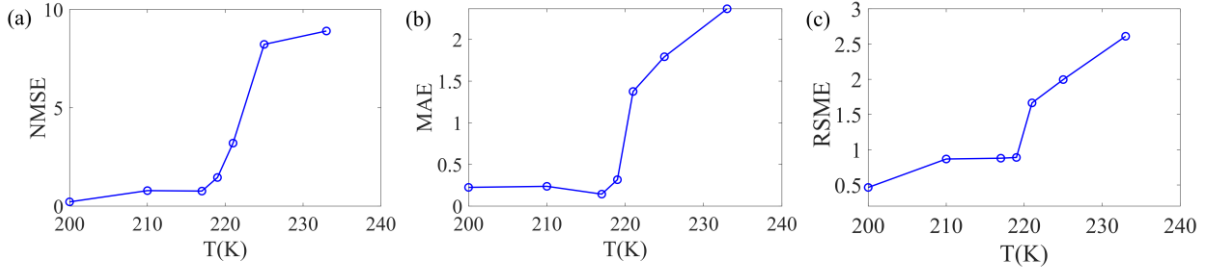

Fig. 10: Variation of the RC metrics $NMSE, MAE$, and $RSME$ as the fuel injection temperature is varied to approach TAI.

The above discussion indicates the potential suitability of the RC metrics $NMSE, MAE$, and $RSME$ for aiding stability analysis and design of a rocket combustor. To be a useful tool for practical use, RC metrics should have a low computational cost. Table 1 shows the cumulative computation time of the RC metrics for a time series length of 0.012 seconds (i.e., the last 20% of the input time series length). Note that the calculation time shown here also includes the training time with the 0.048 seconds-long time series (i.e., 80% of the input time series length). Because the reservoir hyperparameters remain fixed, the RC does not require the input of a human expert. Hence, it can be easily implemented. Based on the above discussion, we infer that the combined LES-RC approach can significantly reduce the computational requirements for the stability analysis of a rocket combustor.

Table 1: Cumulative calculation time of the RC metrics $NMSE, MAE$, and $RSME$ for varying fuel injection temperature. The calculation time shown here includes the training time with the 0.048 seconds-long time series (i.e., 80% of the input time series length) and the computation time of the RC metrics with the 0.012 seconds-long time series (i.e., 20% of the input time series length). In comparison, each LES run requires a computational time of approximately 240 CPU hours using an HPC with 560 Intel Broadwell cores).

| Temperature (K) | 200 | 210 | 217 | 219 | 221 | 225 | 233 |
|---|---|---|---|---|---|---|---|
| Calculation time (seconds) | 0.43 | 0.38 | 0.49 | 0.42 | 0.33 | 0.33 | 0.36 |

**Future scope**

Future works may include exploring the efficacy of the RC approach in mapping the stable regime for various combustor configurations. The efficacy of the RC approach in early-detecting lean blowout in gas turbine combustors can also be studied. Moreover, the physical-substrate-based reservoir computing [75] can be utilized for RC analysis. This will avoid the creation of the simulated reservoir, and computational costs will be reduced further.

**Conclusion**

The present study explores various data-driven and physics-driven techniques that can significantly reduce the computational requirements for the stability analysis of a multi-element, full-scale, supercritical $LO_X$-methane combustor. Pressure time series from the combustor are generated using LES/FGM methodology at seven operating points. We investigate the transition to thermoacoustic instability using three well-known data-driven and physics-driven tools—multi-scale permutation entropy analysis (MPEA), recurrence network analysis (RNA), and reservoir computing (RC) using the pressure time series thus obtained. To classify the stable and unstable combustion regimes, we use the following thresholds for the RC metrics $NMSE \leq 1.5, MAE \leq 0.32,$ and $RSME \leq 0.9$. The results indicate that the combined LES-RC approach is suitable for accurately classifying the stable and unstable combustion regimes of the $LO_X$-methane combustor with only a few data points. Further, the computational requirement for calculating the RC metrics is much less (~ 0.5 seconds using an i7-4770 CPU with 16 GB RAM) as compared to the LES/FGM simulation (~240 CPU hours using an HPC with 560 Intel Broadwell cores). Therefore, the LES-RC approach may be a suitable technique to significantly reduce the computational requirement for stability analysis of the $LO_X$-methane combustor.

**ACKNOWLEDGEMENTS**

The first author would like to thank the Indian Institute of Technology Kanpur (IITK) for the financial support to carry out this work. Also, the authors acknowledge the support of the IITK computer center ([www.iitk.ac.in/cc](www.iitk.ac.in/cc) ) for providing the resources to carry out this work.

**AUTHOR DECLARATIONS**

**CONFLICT OF INTEREST**

The authors have no conflicts to disclose.

**DATA AVAILABILITY**

The data that support the findings of this study are available from the corresponding author upon reasonable request.

**Appendix**

Here, we briefly discuss the method adopted for the construction of the recurrence network. More details can be found in previous studies [26,72]. First, the pressure time series $y(t)$ obtained from CFD simulation is embedded using an optimal embedding dimension ($M$) and optimal time delay ($T_d$). Then, the resulting phase-space vertices are given by,

$$\mathbf{X}(t_i) = [y(t_i), y(t_i + T_d), y(t_i + 2T_d), \dots, \dots, y(t_i + (M-1)T_d)] \qquad (A1)$$

Here, $y(t_i)$ is the value of $\mathbf{y}(t)$ at time instant $t_i$ and $\mathbf{X}(t_i)$ is the corresponding phase-space vector. The recurrence threshold ($\epsilon$) used in this study is calculated to be 10% as per Eroglu et al.'s method [71]. Next, $M$-dimensional hyperspheres with radius $\epsilon$ are drawn around the phase-space vectors. If the hyperspheres are drawn around phase space vectors $\mathbf{X}_i$ and $\mathbf{X}_j$ overlap with each other, then $i$ and $j$ are considered to be connected through a link, and the corresponding element in the recurrence matrix is given by, $\mathbf{R}_{i,j} = 1$. If the hyperspheres do not overlap, $\mathbf{R}_{i,j} = 0$. From the recurrence matrix $\mathbf{R}$, the adjacency matrix $\mathbf{a}$ is calculated as follows,

$$\mathbf{a}_{i,j} = \mathbf{R}_{i,j} - \delta_{i,j}$$

Here, $\delta_{i,j}$ is Kronecker delta. Next, we describe the RN parameters in short.

**Global efficiency ($\eta$):** It is the average efficiency of information transfer in the network and is given by,

$$\eta = \frac{1}{N(N-1)} \sum_{i,j=1, i \neq j}^{N} \frac{1}{l_{i,j}}$$

Here, $N$ is the number of nodes in the network, $l_{i,j}$ is the shortest path length between the nodes $i$ and $j$ [26]. It is calculated with the breadth-first search algorithm [95]. $\eta$ is higher if the dynamical state is deterministic and is lower if the dynamical state is chaotic [26].

**Average degree centrality ($k$):** The degree of a node is given by the number of other nodes it is connected to. The average degree of a network is the average of degrees of all nodes in the network and is given by,

$$k = \frac{1}{N} \sum_{i=1}^{N} k_i$$

Where, $k_i$ is the degree of node $i$ and is given by,

$$k_i = \sum_{j=1, i \neq j}^{N} \mathbf{a}_{i,j}$$

**Global clustering coefficient ($C_G$):** Global clustering coefficient indicates the propensity of the network to form dense interconnected clusters (called a clique). It is given by,

$$C_G = \frac{1}{N} \sum_{i=1}^{N} \left( \frac{e_i}{k_i(k_i-1)/2} \right) = \frac{1}{N} \sum_{i=1}^{N} \left( \frac{2 \sum_{j,v} \mathbf{a}_{i,j} \mathbf{a}_{j,v} \mathbf{a}_{v,i}}{k_i(k_i-1)} \right)$$

Here, $e_i$ indicates the number of nodes directly connected to node $i$. $j$ and $v$ indicate various nodes the node $i$ is connected to.